\documentclass[12pt,a4paper]{article}

\usepackage[british]{babel}

\usepackage[a4paper,top=2cm,bottom=2cm,left=2.5cm,right=2.5cm,marginparwidth=1.75cm]{geometry}

\usepackage[style=apa, backend=biber]{biblatex} 
\DeclareLanguageMapping{british}{british-apa} 
\DeclareFieldFormat[article]{volume}{\apanum{#1}} 

\usepackage{amsmath}
\usepackage{graphicx}
\usepackage[colorlinks=true, allcolors=blue]{hyperref}
\usepackage{hyperref}
\usepackage{orcidlink}
\usepackage[title]{appendix}
\usepackage{mathrsfs}
\usepackage{amsfonts}
\usepackage{booktabs} 
\usepackage{caption}  
\usepackage{threeparttable} 
\usepackage{algorithm}
\usepackage{algorithmicx}
\usepackage{algpseudocode}
\usepackage{listings}
\usepackage{enumitem}
\usepackage{chngcntr}
\usepackage{booktabs}
\usepackage{lipsum}
\usepackage{subcaption}
\usepackage{authblk}
\usepackage[T1]{fontenc}    
\usepackage{csquotes}       
\usepackage{diagbox}
\usepackage{xcolor}
\usepackage{threeparttable}

\definecolor{mygreen}{RGB}{0, 140, 0}
\definecolor{myred}{RGB}{255, 0, 0}

\usepackage{setspace}
\usepackage{titlesec}
\titleformat{\section} 
  {\normalfont\Large\bfseries}{\thesection.}{1em}{}
  
\usepackage{float}   
\usepackage{caption} 


\title{Estimating causal effects of customer satisfaction on downstream metrics in a multi-queue contact center}

\author[1]{Sebastián Orellana}
\author[1]{Leandro Magga}
\author[2,3,4]{Paolo Gorgi}
\author[4]{Hyeokmoon Kweon}
\author[1]{Felipe Bahamonde}

\affil[1]{\small LATAM Airlines, Santiago, Chile}
\affil[2]{\small Vrije Universiteit Amsterdam, Amsterdam, The Netherlands}
\affil[3]{\small Tinbergen Institute, Amsterdam, The Netherlands}
\affil[4]{\small acmetric, Amsterdam, The Netherlands}

\date{}  

\begin{document}
\maketitle


\begin{abstract}

   Contact centers are crucial in shaping customer experience, especially in industries like airlines where they significantly influence brand perception and satisfaction. Despite their importance, the effect of contact center improvements on business metrics remains uncertain, complicating investment decisions and often leading to insufficient resource allocation. This paper employs an instrumental-variable approach to estimate the causal effect of customer service interactions at the contact center of LATAM airlines on downstream metrics. Leveraging observational data and the examiner design, we identify causal effects through the quasi-random assignment of agents to calls, accounting for the multi-queue structure and agent certification heterogeneity. Our empirical results highlight the necessity of an instrumental variable approach to accurately estimate causal effects in contact centers, revealing substantial biases from spurious correlations. This methodology provides managers with tools to estimate the impact of call satisfaction on key business metrics, offering valuable insights to solve operational trade-offs of call centers.
   
\end{abstract}



\section{Introduction}

Call centers are a crucial touchpoint between companies and customers, providing them with support, and significantly shaping their brand perception. In the airline industry, customer service interactions are significant predictors of customer satisfaction and loyalty \parencite{Rhoades2008}. For instance, during service disruptions, such as cancellations and baggage issues, the ability of employees to diagnose and address the situation has shown critical in mitigating the negative effects of the service failure \parencite{Ling2010}. This becomes important in the highly competitive airline industry, where customer dissatisfaction may result in switching to the competition. 


Despite the aforementioned importance of call centers, the broader impact of initiatives aiming at improving the customer experience on business outcomes remains unclear, complicating investment decisions. Conventional methods that only examine correlational relationships could lead to biased estimates as the declared satisfaction during the survey is correlated with unobserved customer characteristics, like customer anger, which act as confounders of the true causal effect. Therefore, relying exclusively on correlational methods may result in inadequate decisions. Uncovering the true causal effect is highly valuable for informed decision-making, offering data-driven insights that enable effective resource allocation and strategic enhancements to boost customer satisfaction.

This paper aims to estimate the causal effects of customer service experience on downstream metrics, such as recontact, satisfaction in future contacts and revenue, using data from an airline company. To achieve this goal, \textit{the examiner design} was applied \parencite{chyn2024examiner}, which considers decision makers who are pseudo-randomly assigned to experimental units and widely differ in some attribute that is correlated with the endogenous variable of interest. For example, \textcite{Stevenson2016} leveraged on the quasi-random assignment of judges in the American justice system to explore the impact of pre-trial detention on case outcomes, and exploited the wide difference in the propensity to set bail between judges to create the instrumental variable, which is a measure of the leniency of each judge. In our context, we exploited the quasi-random variation in the agent availability as an instrument for the service experience leveraging the call center dynamics as a natural experiment. 

The examiner design has been applied in numerous situations to identify causal effects, including the impact of incarceration on criminal behavior and labor market activity \parencite{Mueller2014}, the effect of Chapter 13 bankruptcy protection on financial health applications \parencite{Dobbie2017}, and the impact of pre-trial detention on case outcomes \parencite{Stevenson2016}. In the call center setup, \textcite{Huang2021} used the agent assignment to calls as an instrument to estimate the causal effect of service satisfaction on customer loyalty in a call center of a credit card provider company. However, their methodology does not account for multi-queue dynamics, where agents may receive calls from multiple service queues based on their certifications, which is common in many call center operations \parencite{Aksin2007}. Because agents with different certifications may have systematically different patterns of availability, the assignment of agents to calls may no longer be fully exogenous. 

We propose a simple yet effective identification strategy that uses the examiner design in the context of call center operation, which can account for the multiple-queue dynamics and ensure the exogeneity of the agent assignments. Specifically, we set up short time windows of equal length and condition on them. Because the agent availability is plausibly random within each time span regardless of the multiple queue dynamics, the agent assignment can be a valid instrument conditional on these time spans. We provide a detailed illustration on this strategy as well as supporting empirical evidence.

Our results highlight that conventional methods may underestimate the causal impact of call center enhancements on metrics such as customer recontact. Using the proposed methodology, we quantified the causal impact of the status change from resolved to non-resolved in the self-reported resolution in a \textit{68\% decrease in the probability of recontact within 24 hours}. On the other hand, OLS underestimated a 21\% decrease in the outcome metric.

The paper contributes to the literature in several aspects. First, we explain how the examiner design can be adapted to the multi-queue structure of call center. Second, in an applied case study, we highlight the underestimation of the impact of call center operations on business outcomes, which can potentially lead to under-investment in agent training. Third, we explain how the estimated effects can be used by decision-makers for cost-benefit analyses and to rank proposed improvements for enhancing customer satisfaction.

The remainder of the paper is structured as follows. Section \ref{sec:agent-call-assignment} provides the background on call center operations and explains how call centers historical data can be leveraged as a natural experiment. Section \ref{sec:data} describes the cross-sectional data from an airline call center and presents descriptive analyses of the main variables. Section \ref{sec:identification} details our identification strategy using instrumental variables and accounting for multi-queue agent availability. Section \ref{sec:results} presents the estimated causal effects, and compares the IV estimates with the naive OLS estimates. Finally, Section \ref{sec:conclusions} discusses the managerial implications and the business value of the causal insights.

\section{Background}
\label{sec:agent-call-assignment}

\subsection{Call center operations as a natural experiment}

This section briefly describes how the assignment of calls to agents operates in the multi-queue and multi-certification call center of the studied airline. This mechanism is crucial for identifying causal effects using the examiner design.

\begin{figure}[!ht]
    \centering
    \includegraphics[scale=0.6]{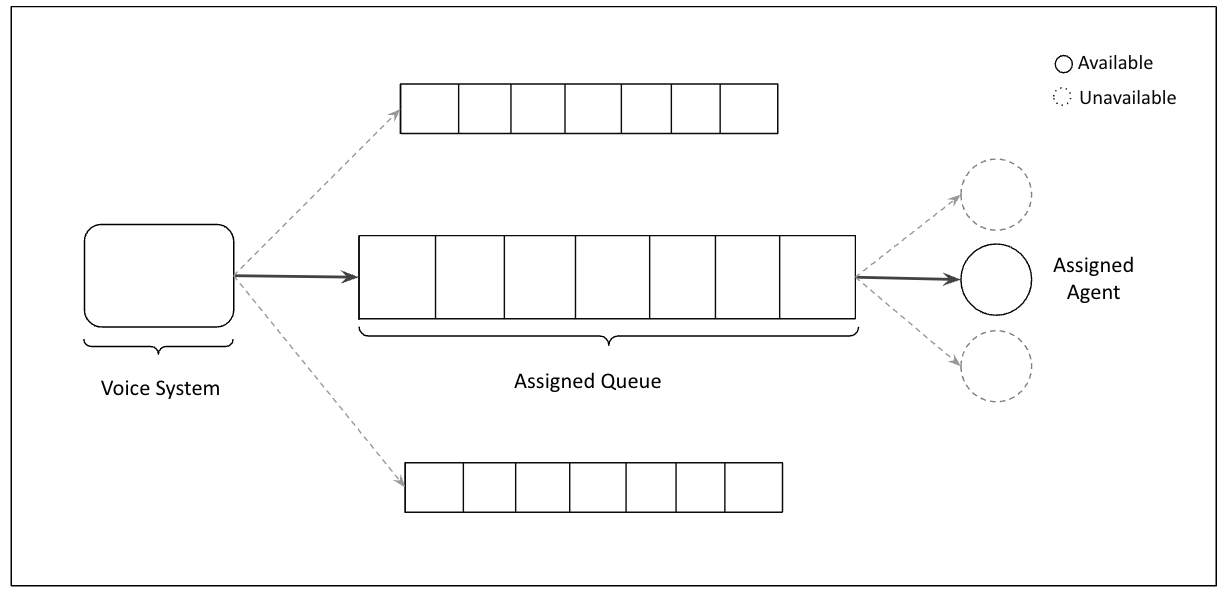}
    \caption{Representation of the agent assignment to calls mechanism based on agents' random variation in availability. The first call in the queue is routed to the first available agent.}
    \label{fig:routing-mechanism}
\end{figure}

The process that each customer follows after calling the airline's contact center is described as follows. First, the customer encounters an interactive voice response system that prompts the customer to choose an option corresponding to their specific requirement. Second, based on the customer's declared issue, the system directs the customer to a specific queue where the customer awaits until an agent with the certification to operate in that queue becomes available. There is no established priority rule among customers; that is, service is provided on a first-come, first-served basis. This agent-call assignment mechanism, depicted in Figure \ref{fig:routing-mechanism}, creates a pseudo-random assignment of examiners that is only based on the agents' random variation in availability, which may be correlated with the timing of the call.

\begin{figure}[ht!]
    \centering
    \includegraphics[scale=0.35]{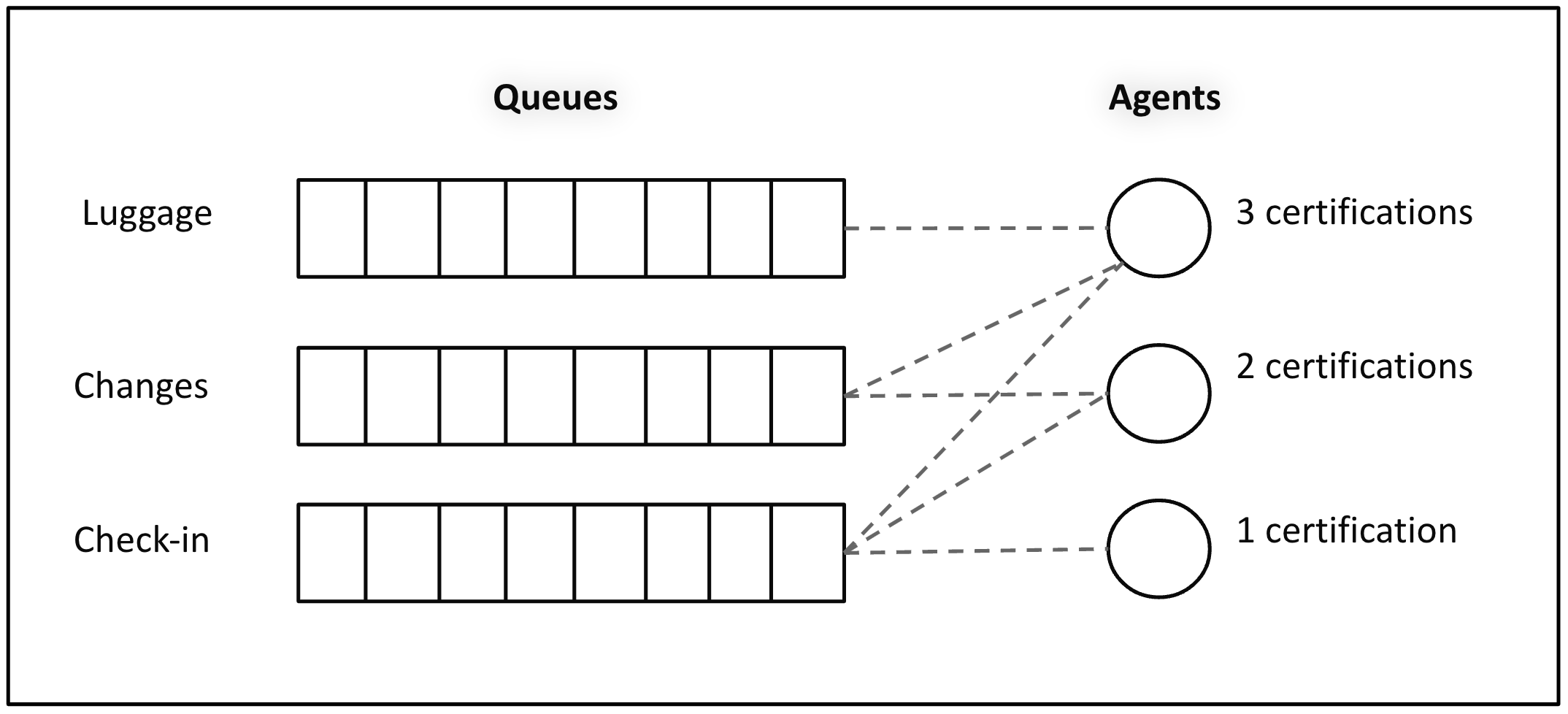}
    \caption{Agent multi-queue availability based on their certifications. In the illustration, the agent at the top has the three certifications, and therefore, may receive calls from any of the three queues.}
    \label{fig:multi-queues}
\end{figure}

Another important feature of call center operations at the airline is the fact that agents are eligible to handle calls from multiple service queues based on certifications. More specifically, agents are classified based on the skills they acquire through certifications, and these certifications determine their active queues on the routing system. As shown in Figure \ref{fig:multi-queues}, there exists heterogeneity in the queues that agents operate. For a given queue, the availability of agents with multiple certifications depends on how busy other queues are and agents with multiple certifications may have different levels of skills compared to agents with less certifications. External factors such as higher demand on complex issues, may temporarily redirect the best agents to other queue and leaving the less skilled agents on one queue. This multi-queue system is fully taken into account in our empirical strategy as will be described below. 


\section{Data}
\label{sec:data}

\subsection{Call center data}

Our data consists of calls received at an airline call center between March 1st, 2023 and May 1st, 2023. For our empirical research, we excluded calls involving \textit{i}) agent transfers and \textit{ii}) calls with unknown customer identification. The first condition is necessary because calls with transfers are not pseudo-randomly received by agents, which would invalidate our identification strategy. The second condition is required to associate customer level downstream metrics with each call as described below. Additionally, we focus on calls received on a specific queue which handles flight changes for Spanish-speaking countries and low-tier customers. This queue is significant because it handles a substantial portion of the airline's call volume, and because it has a diverse range of agent skill levels. The final dataset includes 65,012 calls related to the aforementioned queue,  with customer identification and anonymized phone numbers from the automatic number identification system.

The Genesys system provided detailed call data, including basic features, time spent on various processes, routing, agent assignment, and customer feedback from the post-attention survey. We integrated key business metrics such as the 24-hour recontact indicator, revenue in the following months, claims, and lawsuits, to assess causal impacts. One aspect in the data processing is to identify related customers and aggregate these business metrics at what we call ``family level''. Taking the transitive closure between customer phones, allowed us to group customers who shared the same phone number when interacting with the call center, or whom are transitively connected. With this strategy, we were able to increase coverage of the calls identification, and therefore, to increase the number of observations in our data set by about 10\%. The aggregation is also relevant because often customers call from different cell phones to the contact center (specially when they had a first call with no resolution) and hence, recontacts from the same issue may occur form different cell phones. Likewise, a relative of the customer (who is also a customer) may call for the same issue regarding the family. Therefore, this level of abstraction allows us to better reflect the way that the recontact works. Finally, this analysis also allowed us to identify and filter calls made by travel agencies which could have introduced noise into our study. 

\subsection{Main metrics for analysis}
We consider recontact withing 24 hours as our main downstream metric to illustrate the approach. This is a binary metric that takes the value of 1 when either a customer's phone number or a customer identification from the same family group makes another call to the call center within 24 hours of the previous interaction. Recontact is commonly interpreted as a strong indicator that the customer's initial issue was not resolved satisfactorily and is aligned with the strategic goal of total resolution at the airline. The choice of this metric is also supported by existing studies which suggest that unresolved issues often lead to repeated contact, which can serve as a proxy for customer dissatisfaction and service failure \parencite{Johnston2008}. Moreover, by focusing on a 24-hour window, we capture a critical timeframe that aligns with the operational goals of minimizing disruptions and ensuring customer satisfaction. 

As measures of the skill level of agents, we collected two self-reported measures: \textit{i}) first contact resolution (FCR) and \textit{ii}) customer satisfaction scores (CSAT). FCR is a binary metric that measures whether the customer issue was solved during their first interaction with an agent. On the other hand, CSAT measures how satisfied the customer is with the service interaction in a scale from 0 to 5. Both metrics are reported by the customer during a post-attention survey at the end of the call and provide insights into customer satisfaction. However, they capture different aspects of the service experience. FCR focuses specifically on the efficiency and effectiveness of issue resolution, and is particularly important for both operational efficiency and customer convenience, as unresolved issues often lead to customer frustration and additional resource utilization \parencite{Feinberg2000, Gans2003}. Conversely, CSAT captures the overall satisfaction of the customer with the service interaction, and encompasses not only issue resolution but also the customer's perception of the agent's behavior, communication skills, and the overall service environment \parencite{deRuyter2002}. Therefore, CSAT provides a broader assessment of the customer's experience, including emotional and qualitative aspects that FCR might not fully capture. Nevertheless, CSAT is bimodal and, because it ranges from 0 to 5, it  variance is higher than the binary FCR.

\subsection{Descriptive statistics}
A summary of the data is presented in Table \ref{tab:descriptive-stats}. The data includes 214,013 calls from the studied queue\footnote{This queue handles ticket changes for Spanish-speaking countries with agents based in Colombia.}, with 81.612 (38\%) of these calls resulting in customers receiving a post-attention survey before ending the call. Additionally, note that the distribution of the customer satisfaction score (our endogenous metric) is bimodal; customers tends to give ratings on the extremes –often either giving a perfect score of 5 when fully satisfied or the lowest score of 0 when dissatisfied. 

\setlength{\tabcolsep}{8pt}
\begin{table}[H] \footnotesize
\centering
\caption{\label{tab:descriptive-stats}Descriptive Statistics for the data in the studied queue.}
\begin{tabular}{lcc}
    \hline \\ [-0.8em]
    \multicolumn{1}{c}{\textbf{ }} & \multicolumn{2}{c}{\textbf{Recontact 24h Indicator}} \\ \cline{2-3} \\ [-0.8em]
    \textbf{Variable} & \textbf{0}, N = 133,943 & \textbf{1}, N = 80,060 \\ 
    \toprule
    Waiting Time & 136.69 (418.00) & 114.01 (362.49) \\ 
    Call Language &  &   \\ 
        \hspace{1em} English & 2,339 (1.7\%) & 1,727 (2.2\%)  \\ 
        \hspace{1em} Spanish & 131,604 (98.3\%) & 78,343 (98\%)  \\ 
    Customer Tier &  &  \\ 
        \hspace{1em} Z & 11 (<0.1\%) & 3 (<0.1\%)  \\ 
        \hspace{1em} C & 173 (0.1\%) & 77 (<0.1\%) \\ 
        \hspace{1em} B & 63 (<0.1\%) & 49 (<0.1\%)\\ 
        \hspace{1em} A & 934 (0.7\%) & 468 (0.6\%)  \\ 
        \hspace{1em} D & 96,295 (72\%) & 55,651 (70\%)   \\ 
        \hspace{1em} No Info & 36,467 (27\%) & 23,822 (30\%)   \\ 
    Call EPA Release & 76,507 (57\%) & 26,395 (33\%)  \\ 
    Call EPA Satisfaction Score &  &  \\ 
        \hspace{1em} 0 & 12,263 (20\%) & 7,096 (34\%)   \\ 
        \hspace{1em} 1 & 2,953 (4.8\%) & 2,681 (13\%)  \\ 
        \hspace{1em} 2 & 759 (1.2\%) & 377 (1.8\%)   \\ 
        \hspace{1em} 3 & 1,419 (2.3\%) & 527 (2.5\%)   \\ 
        \hspace{1em} 4 & 4,267 (7.0\%) & 1,144 (5.5\%)   \\ 
        \hspace{1em} 5 & 39,327 (64\%) & 8,885 (43\%)   \\ 
    Call EPA Resolution Score & 44,817 (73\%) & 9,972 (48\%) \\ 
    \bottomrule
\end{tabular}
\end{table}

\section{Identification strategy}
\label{sec:identification}

\subsection{Overview}

Our objective is to estimate the causal effect of costumer satisfaction $Sat$ on an outcome metric $Y$, such as customer recontact within 24 hours. We consider the following model to describe the outcome metric as a function of service satisfaction and other exogenous variables:

\begin{equation}
\label{main_eq}
    Y_{i} = \beta_0 + \beta_1   Sat_{i} + \theta'W_{i} + \varepsilon_{i},
\end{equation}

\noindent where $i=1,\dots, n$ is the index that identifies the call, $Sat_{i}$ represents the reported satisfaction by the customer for the call (FCR or CSAT), $Y_{i}$ is the outcome metric observed for the customer of call $i$, $W_{i}$ is a vector of covariate variables, which will be described below, and $\varepsilon_{i}$ is the error term. 

The key challenge is the potential endogeneity of $Sat_{i}$, \textit{i.e.,} the correlation between $Sat_{i}$ and $\varepsilon_{i}$, which can introduce a bias in the standard OLS estimator of the desired causal effect $\beta_1$. For instance, the declared satisfaction may be correlated with customer characteristics such as the customer's leniency with the issue, which can confound the true causal effect.

To identify the causal effect of the customer satisfaction, we exploited the variation in the customer satisfaction due to the agent assignments to calls. In other words, we isolated the between-agent variation in the customer satisfaction score and used it as an instrument, which can reflect the skill level of each agent related to customer satisfaction. Following the standard procedure in the examiner design literature, we constructed the instrument in the jackknife fashion \parencite{arnold2018racial, cunningham2024adverse}. We first regressed the satisfaction score on the covariates
\begin{equation}
    \label{eq:time-effects}
    Sat_{i} = \delta_0 + \delta_1 W_{i} + u_i.
\end{equation}

\noindent We then residualized the score given the estimates from the model above: 

\begin{equation}
    \label{eq:residuals-computation}
    Sat_{i}^{*} = Sat_i - \hat{\delta_0} - \hat{\delta_1} W_{i}.
\end{equation}

\noindent Next, we computed the instrument as the average of the residualized satisfaction scores, excluding the score of the current call as follows:

\begin{equation}
    \label{eq:instrument-computation}
    Z_{i} = \frac{1}{|C_{a(i)}|} \sum_{j\in C_{a(i)}} Sat_{i}^{*},
\end{equation}

\noindent where $C_{a(i)}=\{j | j \neq i, a(j)=a(i))\}$ is the set of calls answered by the agent who handled the $i$-th service call, excluding the $i$-th call. We'll refer to this specification of the instrumental variable as the residualized leave-one-out satisfaction agent score or Agent's LOO Mean Sat. Score in the upcoming sections.

\subsection{Accounting for multiple queues}
As discussed in the previous section, the routing mechanism of agents follows a quasi-random assignment mechanism. However, for a given call, this quasi-random assignment of agents occurs only among agents that are available at the time of the call. Therefore, the validity of using the agent assignment as an instrument requires that the availability of agents over time does not correlate with the unobserved features of the call. However, this assumption is likely to be violated in practice, especially in a multi-queue call center where agents are assigned to different service queues according to their certifications.

\begin{figure}[h!]
    \centering
    \includegraphics[scale=0.5]{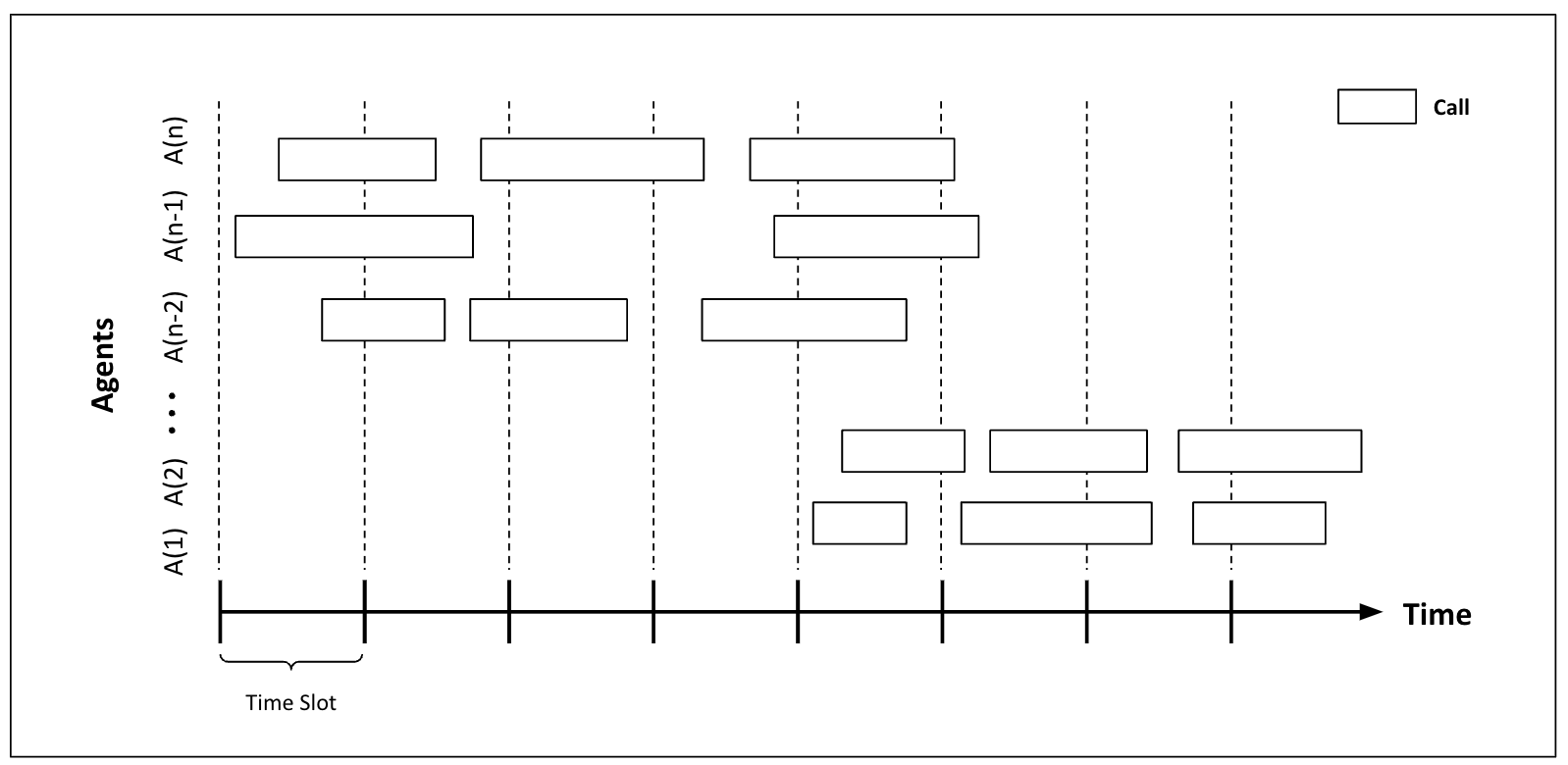}
    \caption{Agent availability and time spans based on the starting time of each call. Note that agents $A(1)$ and $A(2)$ handle calls during similar time periods, possibly due to shift patterns or because they are required in other queues due to their specific certifications. Additionally, agent $A(n-1)$ did not take any calls during the third time slot. This absence could be attributed to scheduled breaks such as lunch or training sessions, or it might simply be a random occurrence where no calls were directed to this agent despite being available.}
    \label{fig:agent-time-spans}
    
\end{figure}

There are a number of reasons that the multi-queue system may invalidate the exogeneity of the agent assignment. For instance, different queues can be more or less busy and the waiting time can correlate with unobserved characteristic of costumers calling. Also, the availability of agents with different sets of certificates in a given queue will differ depending on how busy the other queues are, as shown in Figure \ref{fig:agent-time-spans}. Furthermore, more challenging calls may be clustered over some specific time periods and less skilled agents may take much longer to handle these calls compared to more skilled agents. This will reduce the availability of less skilled agents in these challenging periods, creating a correlation between the availability of agents and unobserved characteristics.

As a simple way to address the issues of the multiple-queue system, we defined arbitrarily short time spans of equal length in which the call started. Then we included a set of time fixed effects as control variables in the model ($W_i$ in equation \ref{main_eq}) that represent these different time spans in which the call was initiated. Specifically, the fixed effect in $W_i$ for a time span $s$ is a dummy variable takes the value 1 if the call $i$ started within the time span $s$ and 0 otherwise. By conditioning on these time fixed effects and focusing on the calls that took place within the same time span, we can remove systematic relationships between the agent availability and the characteristics of the call such as waiting times, agent congestion, type of costumers and the nature of calls.

One then needs to determine the length of each time span. If the span is not short enough, we may end up insufficiently accounting for the multiple queue issue. On the other hand, the model will be overfitting if the time span is too narrow with only a few calls belonging to each window. If this is the case, there may be too little remaining variation in the satisfaction score for the instrument to explain, which will make the instrument weaker. For our main results, we set it to 20 minutes, which results in 12.182 windows containing 17.56 calls on average. In the robustness checks, we present results with different time window lengths.


\subsection{First stage strength}
We examine the relevance of the instrument by examining its ability to explain the variation in the first-stage model, which can be described as below:
\begin{equation}
\label{main_eq}
    Sat_{i} = \gamma_0 + \gamma_1   Z_{i} + \phi'W_{i} + \varepsilon_{i}.
\end{equation}

\noindent Here $\gamma_1$ measures the strength of the instrument $Z_{i}$ conditional on the covariates $W_{i}$. 

Table \ref{tab:first-stage} reports that the instrument has a strong first-stage strength, even after conditioning on the large set of time fixed effects. The effective F statistics are sufficiently high, which exceed 80 for both scores.

\begin{figure}[h!]
    \centering
    \includegraphics[scale=0.6]{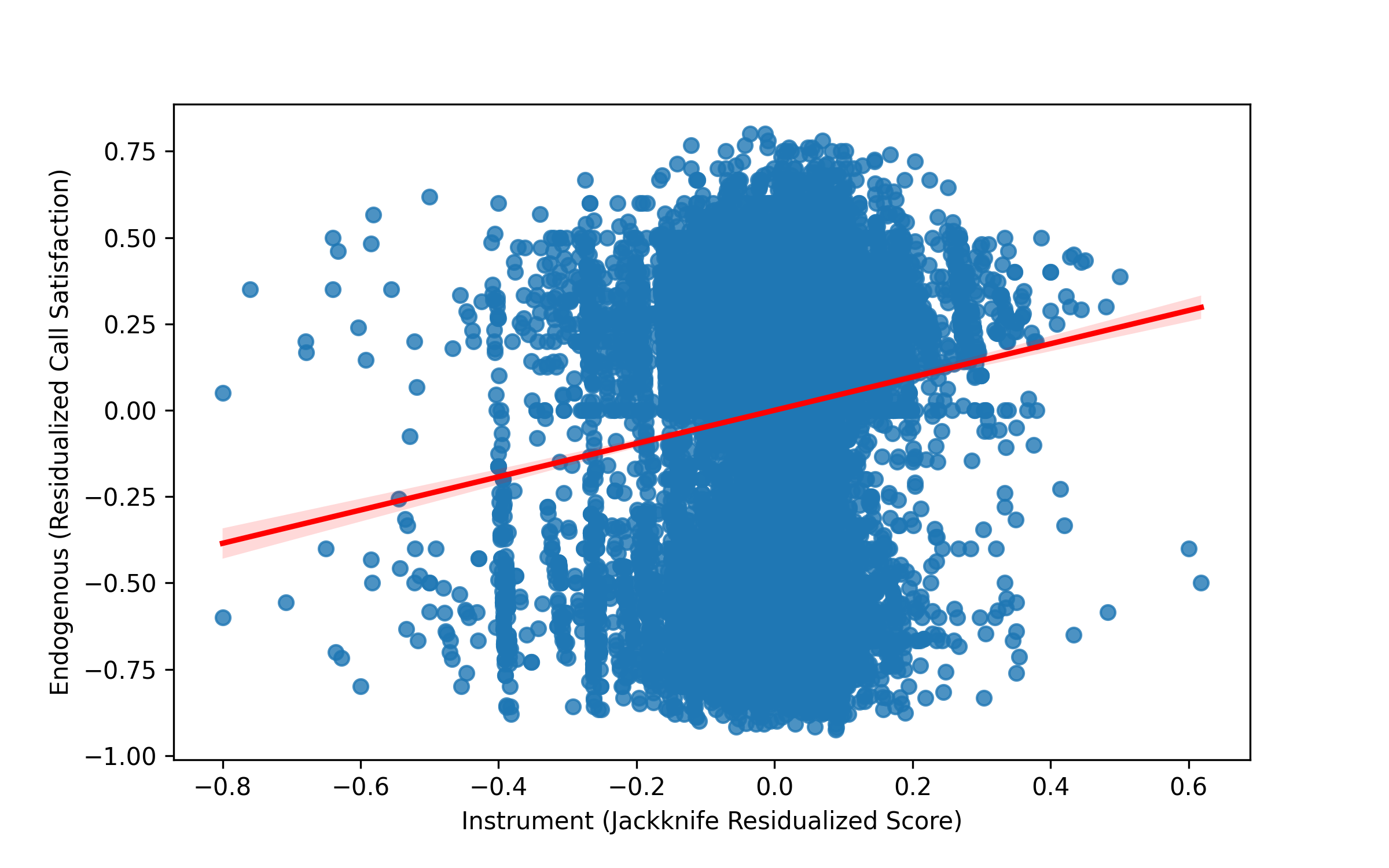}
    \caption{Scatterplot between the instrumental variable (residualized leave-one-out satisfaction agent score) and the endogenous variable (customer satisfaction score with the service call).}
    \label{fig:scatterplot}
\end{figure}

\begin{table*}[h!]
    \centering
    \begin{threeparttable}
    \vspace{1.0em}
    \caption{\label{tab:first-stage}First Stage Regressions for Call Experience Measures.}
    \vspace{-0.9em}
    \small
\begin{tabular}{lcccc}
    \toprule
    \multicolumn{1}{c}{\textbf{ }} & \multicolumn{2}{c}{Call CSAT} & \multicolumn{2}{c}{Call FCR} \\ 
    \cmidrule(lr){2-3} \cmidrule(lr){4-5} \\ [-0.8em]
                                        & \text{(1)}      & \text{(2)}   & \text{(3)}     & \text{(4)} \\
    \midrule
    Z: Agent's LOO Mean Sat. Score       & $0.632^{***}$   & $0.634^{***}$  & $0.630^{***}$       & $0.630^{***}$ \\
                                        &  $(0.033)$      & $(0.033)$      & $(0.036)$             & $(0.036)$  \\
    \midrule
    Kleibergen-Paap F-Test              & 365.57            & 364.57          & 310.95            & 310.69   \\
    Time Controls                       & Yes            & Yes          & Yes           & Yes  \\
    Baseline Controls                   & Yes            & No           & Yes           & No  \\
    Observations                        & 81,612         & 81,612       & 81,612        & 81,612 \\
    \bottomrule
\end{tabular}
\begin{minipage}{13cm}
\vspace{0.1cm} \footnotesize
    Notes: Standard errors are computed using a heteroskedasticity and autocorrelation correction and the robust (Kleibergen-Paap) first stage F is reported. Note, this is equivalent to the effective F-statistic of \textcite{OleaJoséLuisMontiel2013ARTf} in this case of a single instrument \parencite{LewisDanielJ2022ARTf}. Significance levels: $^{*} p<0.10$, $^{**} p<0.05$, $^{***} p<0.01$.
\end{minipage}
    \end{threeparttable}
\end{table*}

\subsection{Instrument validity}

In addition to the relevance, a valid instrument needs to satisfy the conditional independence. In other words, the instrument assignment must be as good as random given the covariates. In our setup, this requires that the agents are assigned to calls randomly. However, as described above, the agent assignment may not be fully random in a multi-queue system. Here we provide supporting evidence that the agent assignment is as good as random within a short time frame and the conditional independence can only hold when the time fixed effects are accounted for. 

First, as one way to show that the agent assignment is not fully random in multiple queues, we examine the relationship between the waiting time of the calls and the agent assignment. Specifically, we regressed the waiting time on the agent fixed effects with or without the time fixed effects. If the agent assignment were to be random, we should expect to observe no systematic difference in the waiting time depending on which agent handled the call. Table \ref{tab:waiting-time-models} clearly shows that the waiting time differs across the agents. The agent fixed effects can jointly account for a substantial amount of the variation in the waiting time ($R^2=83.4\%$). This relationship is statistically significant based on the F-test (\textit{p-value} $<0.001$). Once the time fixed effects are controlled for, the agent fixed effects only explains around 1\% of the variation in the waiting time, which is statistically not significant (\textit{p-value} $= 0.944$), supporting that the agent assignment is plausibly random conditional on time fixed effects. 

Second, we perform randomization tests to evaluate the conditional independence of our instrument $Z_i$. Table \ref{tab:randTest} provides supporting evidence that the conditional independence only holds conditional on the time fixed effects. The first column shows that the moderate correlation exists between a call satisfaction score and a set of call characteristics (the joint \textit{F-test p-value} $<0.001$). This correlation is still present after conditioning on the time fixed effects, as reported in the third column (\textit{F-test p-value} $<0.001$ ). In the second column, we conducted the same analysis using the leave-one-out score without considering the time fixed effects. The results do not provide sufficient support for the random assignment of calls when we do not account for the time fixed effects and hence the multi-queue system (\textit{F-test p-value} $<0.001$). Finally, the fourth column reports the results fully taking into account the fixed effects. As shown, we no longer observe the correlation between the the leave-one-out score and the call characteristics (\textit{F-test p-value} $= 0.44$), supporting that the conditional independence holds in this setup.

\begin{table*}[h!]
    \centering
    \begin{threeparttable}
    \vspace{1.0em}
    \caption{\label{tab:waiting-time-models}Conditional independence assessment via waiting time models.}
    \vspace{-0.9em}
    \small
\begin{tabular}{lcc}

\toprule
                     & \multicolumn{2}{c}{Dependant: Waiting Time} \\
    \cmidrule(lr){2-3} \\ [-0.8em]
                     &        (1)            &        (2)              \\
    \midrule
    Agent's Coefficients Joint F-Test   &   $26.73^{***}$        &   $0.283$              \\
    \midrule
    Time Controls    &        No            &        Yes               \\
    Model $R^{2}$          &   $0.834$             &   $0.844$              \\
     Net \%  variation explained by the Agent Dummies    &          $0.834$             &   $0.01$              \\
\bottomrule
\end{tabular}

\begin{minipage}{14cm}
\vspace{0.1cm} \footnotesize
    Notes: This table reports the joint hypotheses F-test that was conducted using the waiting time models. The significance of the coefficients associated with the agents dummies indicates that time controls are relevant for the conditional independence assumption. Standard errors are using a heteroskedasticity and autocorrelation correction. Significance levels: $^{*} p<0.10$, $^{**} p<0.05$, $^{***} p<0.01$.
\end{minipage}
    \end{threeparttable}
\end{table*}

\begin{table}[htpb!]
     \centering
     \begin{threeparttable}
     \vspace{1.0em}
     \caption{\label{tab:randTest}Test of randomization.}
     \vspace{-0.9em}
     \footnotesize
\begin{tabular}{lcccc}
\toprule                                              
& 
\begin{tabular}[c]{@{}c@{}}(1)\\Call\\ CSAT\end{tabular} & 
\begin{tabular}[c]{@{}c@{}}(2)\\Z: Agent LOO\\ CSAT\end{tabular} & 
\begin{tabular}[c]{@{}c@{}}(3)\\ Call\\ CSAT\end{tabular} & 
\begin{tabular}[c]{@{}c@{}}(4)\\ Z: Agent LOO\\ CSAT\end{tabular}\\\midrule
Call Market             &         &         &         &            \\
\hspace{2em} AU         & -0.011  & -0.021   & -0.069  &  -0.040     \\
                        & (0.071) & (0.014) & (0.075) & (0.014)    \\
\hspace{2em} BR         & -0.086  & 0.010   & -0.071   &  0.014     \\
                        & (0.064) & (0.009) & (0.064) & (0.010)    \\
\hspace{2em} CL         & -0.001   & -0.002   & 0.002  &  -0.001     \\
                        & (0.006) & (0.002) & (0.006) & (0.001)    \\
\hspace{2em} CO         & 0.038   & -0.002   & 0.039  &  -0.001     \\
                        & (0.006) & (0.001) & (0.006) & (0.001)    \\
\hspace{2em} EC         & 0.024   & -0.002   & 0.047  &  -0.002     \\
                        & (0.023) & (0.005) & (0.024) & (0.005)    \\
\hspace{2em} EU         & -0.045  & -0.001   & -0.063  &  -0.006     \\
                        & (0.021) & (0.005) & (0.025) & (0.005)    \\
\hspace{2em} MX         & 0.011   & -0.002   & 0.030  &  -0.000     \\
                        & (0.036) & (0.008) & (0.038) & (0.008)    \\
\hspace{2em} PE         & 0.022   & -0.002   & 0.017  &  -0.001     \\
                        & (0.007) & (0.002) & (0.008) & (0.001)    \\
\hspace{2em} UK         & -0.159  & -0.043   & -0.219  &  -0.084     \\
                        & (0.132) & (0.022) & (0.160) & (0.027)    \\
\hspace{2em} US         & -0.042  & -0.031   &  -0.078 &  -0.047     \\
                        & (0.076) & (0.015) & (0.085) & (0.016)    \\
\hspace{2em} UY         & -0.006  & 0.003   & -0.012   &  0.008     \\
                        & (0.030) & (0.005) & (0.032) & (0.005)    \\
\hspace{2em} ES         & -0.054  & -0.032   & -0.093  &  -0.049     \\
                        & (0.074) & (0.014) & (0.082) & (0.016)    \\
FFP Category            &         &         &         &            \\
\hspace{2em} C   & -0.147  & 0.036   & -0.136   &  0.036  \\
                        & (0.216) & (0.017) & (0.243) & (0.019) \\
\hspace{2em} B  & -0.120  & 0.005   & -0.124   &  0.007  \\
                        & (0.229) & (0.019) & (0.254) & (0.022) \\
\hspace{2em} A       & -0.039  & 0.029   & -0.013   &  0.035  \\
                        & (0.209) & (0.015) & (0.234) & (0.018) \\
\hspace{2em} D      & -0.067  & 0.028   & -0.052   &  0.033  \\
                        & (0.206) & (0.014) & (0.232) & (0.017) \\
\hspace{2em} NoInfo     & -0.077  & 0.028   & -0.060   &  0.033  \\
                        & (0.206) & (0.014) & (0.232) & (0.017) \\
Log Hours From Last Call& 0.000  & -0.000   & 0.000  & -0.000    \\
                        & (0.001) & (0.000) & (0.001) & (0.000) \\
Bookings Past 12 Months & -0.001  & 0.000   & -0.001   &  0.000  \\
                        & (0.000) & (0.000) & (0.000) & (0.000) \\ \midrule
Time fixed effects      & No      & No      & Yes     &  Yes    \\
F-test                  & 6.32       & 1.048       & 1.091       &  1.016      \\
Observations            & 81,698       & 81,612       & 81,698       &  81,162      \\

\bottomrule
\end{tabular}   

\begin{minipage}{13cm}
\vspace{0.1cm} \footnotesize
    Notes: These linear models control for the baseline characteristics used in the instrumental variables analyses. The dependent variable in columns (1) and (3) is the normalized call CSAT score recontact within 24 hours. The dependent variable in columns (2) and (4) is the propensity to obtain high or low score by agents. Time fixed effects include time control effects $W_t$. Standard errors shown in parentheses. *p<0.10, **p<0.05, *** p<0.01.
\end{minipage}
     \end{threeparttable}
\end{table}

\section{Results}
\label{sec:results}
\subsection{Main results}

In this section we present evidence examining the causal effects of the call satisfaction score on call outcomes. The regressions in Table \ref{tab:estimated-relationship-epa} show the estimated relationship between customer self-reported resolution and customer recontact within 24 hours. Columns (1) and (2) present both the OLS estimate and the 2SLS estimates with the agent leave-one-out average satisfaction as the instrumental for the customer self-reported satisfaction. Note that in both cases we are controlling by the time-dummies $W_{t}$ and hence complying with the conditional independence assumption and by extension the exclusion restriction. When comparing columns (1) and (2) it is clear that OLS underestimates the causal impact of resolution on recontact probability estimating a 22 percentage point decrease due to self-reported call resolution state. Similarly, the 2SLS estimate of the status change from resolved to non-resolved in the self-reported resolution is a \textit{65 percentage point decrease in the probability of recontact within 24 hours.} Moreover, columns (3) and (4) in Table \ref{tab:estimated-relationship-epa} show the estimated relationship between customer self-reported satisfaction and recontact within 24 hours. Similarly as with resolution we observe that OLS underestimates the impact satisfaction on recontact probability estimating a 22 percentage point decrease due to a 5 points increase in self-reported call satisfaction, on the other hand a 5 points increase in self-reported satisfaction leads to a \textit{69 percentage point decrease in the probability of recontact within 24 hours.} Time controls $W_t$ and baseline covariates are included in each of the regressions as it was discussed on previous sections.

\begin{table*}[h!]
    \centering
    \begin{threeparttable}
    \vspace{1.0em}
    \caption{\label{tab:estimated-relationship-epa}Estimated relationship between customer self-reported metrics and recontact within 24 hours using the proposed IV methodology compared to the biased OLS estimate.}
    \vspace{-0.9em}
    \small
\begin{tabular}{lcccc}
    \toprule
    \multicolumn{1}{c}{} & \multicolumn{4}{c}{Outcome: Recontact 24hrs} \\ \cline{2-5} \\ [-0.8em]
                      &        (1)          &        (2)              &        (3)          &        (4)          \\
                      &        OLS          &        2SLS             &        OLS          &        2SLS         \\
     \midrule
     Call FCR         &     $-0.2174^{***}$   &       $-0.6562^{***}$   &                     &        \\
                      &       (0.0066)        &       (0.0403)          &                     &                   \\
     Call CSAT        &                      &                        &  $-0.2073^{***}$    &      $-0.6949^{***}$    \\
                      &                      &                        &      (0.007)       &       (0.0469)            \\
     \midrule
     Time Controls    &        Yes          &        Yes              &        Yes          &        Yes                \\
     Observations     &        81,698        &        81,612            &        81,698        &      81,612              \\ 
     \midrule
     Instrument       &         –           &  Agent LOO FCR          &        –            &  Agent LOO CSAT           \\
     F-Test           &          –          &        310.69               &         –       &       364.57                 \\   
    \bottomrule
\end{tabular}

\begin{minipage}{15cm}
\vspace{0.1cm} \footnotesize
    Notes: This table reports the ordinary least squares (OLS) and the instrumental variables regression (2SLS) estimates of the impact of the call experience metrics on customer recontact within 24 hours. The 2SLS instruments are the agent skill proxies built as using the leave-one-out after residualizing using the time spans. We included the time spans controls in each model. Standard errors are clustered by agent and time period. Significance levels: $^{*} p<0.10$, $^{**} p<0.05$, $^{***} p<0.01$.
\end{minipage}
    \end{threeparttable}
\end{table*}

\subsection{Alternative metrics}

Alternatively we explored the impact of self-reported satisfaction on several other operational metrics, namely next-call self-reported satisfaction, post-call claims within 7 and 28 days from a given call, refund request within a month from a given call,  post-call claims made to a regulatory entity and claims flagged with a high priority.

 For next-call self-reported satisfaction and post-call claims within 7 days the results are presented in Table \ref{tab:estimated-relationship-epa-csat-next-call} which first shows the estimated relationship between customer self-reported satisfaction and next-call self-reported satisfaction. It is important to emphasize that the sample for which we have the recorded next-call self-reported satisfaction is smaller compared with the original sample, consisting of only 19,988 observations for the whole study period. By inspecting columns (1) and (2) we note that OLS overestimates the relationship between consecutive reported satisfaction scores detecting a positive effect of self-reported satisfaction on the average satisfaction reported in the next interaction of 17 percentage points. However when estimating the same effect using the 2SLS identification the detected effect becomes non-significant, suggesting the original positive effect is only the results of uncontrolled confounders in the OLS setting. Similarly, columns (3) and (4) display the relationship between customer self-reported satisfaction and claim submission within 7 days from the original contact center interaction. OLS estimates a small negative but significant effect of 1.8 percentage points for a 5 points increase in self-reported satisfaction and claim submission; however, when the same effect is estimated using a 2SLS identification strategy the causal estimation becomes non-significant again suggesting that OLS is not a suitable identification strategy given the omitted variable bias affecting its estimates.

\begin{table*}[h!]
    \centering
    \begin{threeparttable}
    \vspace{1.0em}
    \caption{\label{tab:estimated-relationship-epa-csat-next-call}Estimated relationship for the studied queue between customer self-reported satisfaction and Next-Call self-reported satisfaction.}
    \vspace{-0.9em}
    \small
\begin{tabular}{lcccc}
    \toprule
    \multicolumn{1}{c}{} & \multicolumn{2}{c}{Next-Call CSAT} & \multicolumn{2}{c}{Claims 7 Days} \\
    \cmidrule(lr){2-3} \cmidrule(lr){4-5} \\ [-0.8em]
                     &        (1)          &        (2)              &        (3)          &        (4)          \\
                     &        OLS          &        2SLS             &        OLS          &        2SLS         \\
     \midrule
     Call CSAT       &     $0.173^{***}$  &    0.0257              &     $-0.018^{***}$  &             -0.007   \\
                     &     (0.009)        &    (0.061)            &      (0.001)        &            (0.009)    \\
     \midrule
     Time Controls    &        Yes          &        Yes             &        Yes          &        Yes        \\
     Observations     &      19,988          &    19,975               &        81,698        &           81,612      \\ 
     \midrule
     Instrument       &         –           &  Agent LOO CSAT        &           –          &  Agent LOO CSAT        \\
     F-Test           &        –            &    364.57                 &           –          &             364.57       \\
    \bottomrule
\end{tabular}
\begin{minipage}{15cm}
\vspace{0.1cm} \footnotesize
    Notes: This table reports the ordinary least squares (OLS) and the instrumental variables regression (2SLS) estimates of the impact of the call experience metrics on next-call CSAT and 7 days claims. The 2SLS instruments are the agent skill proxies built as using the leave-one-out after residualizing using the time spans. We included the time spans controls in each model. Standard errors are clustered by agent and time period. Significance levels: $^{*} p<0.10$, $^{**} p<0.05$, $^{***} p<0.01$.
\end{minipage}
    \end{threeparttable}
\end{table*}

The regressions in Table \ref{tab:estimated-relationship-epa-csat-refund-request} show the relationship between the Refund Request Flag and post-call claims within 28 days and customer self-reported satisfaction. By inspecting columns (1) and (2) we note that there is almost no effect on self-reported satisfaction score with refund requests, with OLS estimating an albeit small but significant effect which disappears when estimating the same effect using the 2SLS identification strategy. On the other hand, columns (3) and (4) display the relationship between customer self-reported satisfaction and claim submission 28 days from the original contact center interaction. OLS estimates a negative effect of 4.8 percentage points for a 5 points increase in self-reported satisfaction on claim submission, however the same effect estimated using the 2SLS identification strategy is non-significant, suggesting potential spurious correlations when using OLS.


\begin{table*}[h!]
    \centering
    \begin{threeparttable}
    \vspace{1.0em}
    \caption{\label{tab:estimated-relationship-epa-csat-refund-request}Estimated relationship for the studied queue between customer self-reported satisfaction and Refund Request and Claim 28 Days flags.}
    \vspace{-0.9em}
    \small
\begin{tabular}{lcccc}
    \toprule
    \multicolumn{1}{c}{} & \multicolumn{2}{c}{Refund Request Flag} & \multicolumn{2}{c}{Claims 28 Days} \\
    \cmidrule(lr){2-3} \cmidrule(lr){4-5} \\ [-0.8em]
                     &        (1)          &        (2)              &        (3)          &        (4)          \\
                     &        OLS          &        2SLS             &        OLS          &        2SLS         \\
     \midrule
     Call CSAT       &     $0.001^{***}$  &    0.000              &     $-0.0487^{***}$    &             -0.0146    \\
                     &     (0.000)        &    (0.000)    &      (0.0022)    &              (0.0167)    \\
     \midrule
     Time Controls    &        Yes          &        Yes             &        Yes          &        Yes        \\
     Observations     &      81,698          &    81,612               &        81,698          &    81,612      \\ 
     \midrule
     Instrument       &         –           &  Agent LOO CSAT        &           –          &  Agent LOO CSAT        \\
     F-Test           &        –            &    364.57                 &           –          &             364.57       \\
    \bottomrule
\end{tabular}

\begin{minipage}{15cm}
\vspace{0.1cm} \footnotesize
    Notes: This table reports the ordinary least squares (OLS) and the instrumental variables regression (2SLS) estimates of the impact of the call experience metrics on Refund Request Flag and 28 days claims. The 2SLS instruments are the agent skill proxies built as using the leave-one-out after residualizing using the time spans. We included the time spans controls in each model. Standard errors are clustered by agent and time period. Significance levels: $^{*} p<0.10$, $^{**} p<0.05$, $^{***} p<0.01$.
\end{minipage}
    \end{threeparttable}
\end{table*}

Finally Table \ref{tab:estimated-relationship-epa-csat-regulatory-claim} shows the relationship between claims submitted to regulatory entities, claims with high internal prioritization, and self-reported satisfaction. By inspecting columns (1) and (2), we note that OLS estimates a small but significant negative effect of 1.1 percentage points on claims flagged as submitted to regulatory entities. However, when comparing with 2SLS, it becomes evident that the OLS estimates may be driven by spurious correlations within the data. Furthermore, columns (3) and (4) display the relationship between customer self-reported satisfaction and claims flagged with high prioritization. Here, OLS results indicate a negative effect of 1.8 percentage points, suggesting that a 5-point increase in self-reported satisfaction is associated with a reduction in high-priority claims, though this effect also appears to stem from spurious correlations.


\begin{table*}[h!]
    \centering
    \begin{threeparttable}
    \vspace{1.0em}
    \caption{\label{tab:estimated-relationship-epa-csat-regulatory-claim}Estimated relationship for the studied queue between customer self-reported satisfaction and Regulatory Entity Claim and High Priority Claim flags.}
    \vspace{-0.9em}
    \small
\begin{tabular}{lcccc}
    \toprule
    \multicolumn{1}{c}{} & \multicolumn{2}{c}{Regulatory Entity Claim Flag} & \multicolumn{2}{c}{High Priority Claim Flag} \\
    \cmidrule(lr){2-3} \cmidrule(lr){4-5} \\ [-0.8em]
                     &        (1)          &        (2)              &        (3)          &        (4)          \\
                     &        OLS          &        2SLS             &        OLS          &        2SLS         \\
     \midrule
     Call CSAT       &      $-0.011^{***}$    &             -0.0055              &     $-0.0182^{***}$    &             -0.0077    \\
                     &     (0.001)    &              (0.0081)    &      (0.0013)    &              (0.009)    \\
     \midrule
     Time Controls    &        Yes          &        Yes             &        Yes          &        Yes        \\
     Observations     &      81,698          &    81,612               &        81,698          &    81,612      \\ 
     \midrule
     Instrument       &         –           &  Agent LOO CSAT        &           –          &  Agent LOO CSAT        \\
     F-Test           &        –            &    364.57                 &           –          &             364.57       \\
    \bottomrule
\end{tabular}

\begin{minipage}{15cm}
\vspace{0.1cm} \footnotesize
    Notes: This table reports the ordinary least squares (OLS) and the instrumental variables regression (2SLS) estimates of the impact of the call experience metrics on Regulatory Entity Claim Flag and High Priority Claim Flag. The 2SLS instruments are the agent skill proxies built as using the leave-one-out after residualizing using the time spans. We included the time spans controls in each model. Standard errors are clustered by agent and time period. Significance levels: $^{*} p<0.10$, $^{**} p<0.05$, $^{***} p<0.01$.
\end{minipage}
    \end{threeparttable}
\end{table*}

\subsection{Robustness checks}

The regressions in Table \ref{tab:estimated-relationship-epa-csat-robust} shows the estimated relationship between customer self-reported satisfaction and customer recontact for different time windows of 48, 72 and 168 hours using the 2SLS estimated as described in the previous section. By comparing columns (1), (2) and (3) we can observe how the estimated causal effect changes when the recontact probability is computed over longer time periods. As expected the effect appears invariant for longer time windows, such as 48 and 72 hours. However there seems to be a slight decay of the effect for the 1 week computation window of the recontact probability (ie: 168 hours), which can be observed in the slight change in the confidence intervals and the magnitude of the treatment effect estimator, which roughly diminishes by 10 percentage points in the case of the self-reported satisfaction.

Moreover the regressions in Table \ref{tab:robustness-wt-length} display the estimated relationship between customer self-reported satisfaction and customer recontact for different lengths of the time controls window computation as described in previous sections. By comparing columns (1) to (5) we can observe how the 2SLS estimator is robust to the selection of the window for the time dummies computation for different specifications. In addition to that we can observe that the selected 20 minutes window is the one which yields the best results in terms of the strength of the Leave-One-Out Mean Satisfaction Score.

\begin{table*}[h!]
    \centering
    \vspace{1.0em}
    \caption{\label{tab:estimated-relationship-epa-csat-robust}Estimated relationship for the studied queue between customer self-reported satisfaction and recontact within 48, 72 and 168 hours}
    \vspace{-0.9em}
    \small
\begin{tabular}{lccc}
\toprule
    Recontact window         & 48 hrs               & 72 hrs              &  168 hrs\\

    \midrule 
     Call CSAT                &  $-0.664^{***}$	    &  $-0.645^{***}$     &   $-0.556^{***}$      \\
                              &  $(0.046)$          &  $(0.048)$          &   $(0.050)$           \\
    \midrule 
    Time Controls             &    Yes              &         Yes         &    Yes              \\ 
    Observations              &    81,612            &         81,612       &   81,612             \\      
    Instrument                &    Agent LOO CSAT   &    Agent LOO CSAT   &    Agent LOO CSAT   \\ 
    \bottomrule
\end{tabular}

\begin{minipage}{15cm}
\vspace{0.1cm} \footnotesize
    Notes: This table reports the instrumental variables regression (2SLS) estimates of the impact of the call experience metrics on Recontact Probability at 48, 72 and 168 hours. The 2SLS instruments are the agent skill proxies built as using the leave-one-out after residualizing using the time spans. We included the time spans controls in each model. Standard errors are clustered by agent and time period. Significance levels: $^{*} p<0.10$, $^{**} p<0.05$, $^{***} p<0.01$.
\end{minipage}
\end{table*}

\begin{table*}[h!]
    \centering
    \vspace{1.0em}
    \caption{\label{tab:robustness-wt-length}Sensitivity Analysis for the length of the $W_t$ time window for the time control dummies, we explored windows of 15, 20, 30, 45 and 60 minutes.}
    \vspace{-0.9em}
    \small
\begin{tabular}{lccccc}
    \toprule
    $W_t$ length in minutes     & 15 min  & 20 min & 30 min  & 45 min & 60 min\\
    \midrule
    Call CSAT                            & $-0.688^{***}$    & $-0.691^{***}$ & $-0.688^{***}$  & $-0.692^{***}$  &  $-0.688^{***}$      \\
                                         & $(0.048)$         & $(0.047)$      & $(0.048)$       & $(0.047)$       & $(0.048)$            \\
    \midrule
    Observations                & 81,612         & 81,612  & 81,612             & 81,612 & 81,612 \\
    Time Controls               & Yes & Yes   & Yes & Yes & Yes\\
    Baseline Controls           & Yes & Yes & Yes & Yes & Yes        \\
    Kleibergen-Paap F-Test      & 269.6 & 373 & 269.8 & 364.5 & 267.6   \\
    \bottomrule
\end{tabular}

\begin{minipage}{15cm}
\vspace{0.1cm} \footnotesize
    Notes: This table reports the instrumental variables regression (2SLS) estimates of the impact of the call experience metrics on Recontact Probability at 24 hours using different length for the time controls of 15, 20, 30, 45 and 60 minutes. The 2SLS instruments are the agent skill proxies built as using the leave-one-out after residualizing using the time spans. We included the time spans controls in each model. Standard errors are clustered by agent and time period. Significance levels: $^{*} p<0.10$, $^{**} p<0.05$, $^{***} p<0.01$.
\end{minipage}
\end{table*}

\section{Conclusion}
\label{sec:conclusions}

Our study highlights how traditional OLS methods might underestimate the causal impact of call center enhancements on downstream metrics such as customer recontact within 24 hours. This biased result could potentially have led to under-investment in this area. Through the proposed IV methodology, we could quantify the true impact from improved service satisfaction empirically which was crucial for providing evidence that justifies increased investment on initiatives aiming at improving agent's training and how agents' balance resolution with productivity.

The estimated causal effects can be used by decision-makers to conduct cost-benefit analyses for proposed improvements that aim to improve self-reported call metrics. This estimated expected impact was used to rank the hypothesis portfolio of initiatives that call center managers use at the airline's call center. Moreover, the analysis was replicated for other queues at the call center, to aid managers to decide where to prioritize initiatives, with the purpose of maximizing the expected value and gaining traction.

Additionally, the estimated causal effect could be used as input for workforce optimization models that optimize the trade-off workforce costs with quality of service. For instance, the model proposed by \textcite{Excoffier2016} that determines the number of agents to be assigned to a set of predefined shifts so as to optimize the trade-off between manpower cost and customer quality of service. The estimated causal effect of customer experience on customer recontact in 24 hours, can be used for matching the best-quality agents to the most complex shifts to decrease the congestion at the call center. 

Finally, the instrumental variable constructed for our research –the residualized leave-one-out agent satisfaction rating– was utilized to developed a ranking system for agents. This allowed managers to study what characterizes the best agents\footnote{Note that this measure informs the satisfaction score received by the agent in the EPA survey compared to other pears in the time slot.} and include new initiatives in the hypothesis portfolio to improve the call center experience.






\printbibliography
\renewcommand\theequation{\Alph{section}\arabic{equation}} 
\counterwithin*{equation}{section} 
\renewcommand\thefigure{\Alph{section}\arabic{figure}} 
\counterwithin*{figure}{section} 
\renewcommand\thetable{\Alph{section}\arabic{table}} 
\counterwithin*{table}{section} 

\begin{appendices}

\end{appendices}

\end{document}